\documentclass[aps,prb,showpacs,twocolumn,superscriptaddress]{revtex4-1}
\usepackage{graphicx,bm,amssymb,amsmath,epsfig,epstopdf}

\newcommand{\Eq}[1]{Eq.~\eqref{#1}}

\newcommand{\beq}{\begin{equation}}
\newcommand{\eeq}{\end{equation}}

\newcommand{\beqa}{\begin{eqnarray}}
\newcommand{\eeqa}{\end{eqnarray}}

\newcommand{\Beqa}{\begin{eqnarray*}}
\newcommand{\Eeqa}{\end{eqnarray*}}
\newcommand{\nn}{\nonumber}

\newcommand{\pdag}{{\phantom{\dagger}}}

\newcommand{\Sinh}[2]{\sinh^{#1}( #2)} 	
\newcommand{\Cosh}[2]{\cosh^{#1}(#2)}	

\newcommand{\arcsinh}[1]{\ensuremath{\mathrm{arcsinh}(#1)}} 

\begin{document}

\title{Signatures of evanescent mode transport in  graphene }

\author{A.D. Wiener}
\author{M.  Kindermann}
\affiliation{ School of Physics, Georgia Institute of Technology, Atlanta, Georgia 30332, USA }

\begin{abstract}
We calculate the shot noise generated by evanescent modes in graphene for several experimental setups.  For two impurity-free graphene strips kept at the Dirac point by gate potentials, separated by a long highly doped region, we find that the Fano factor takes the universal value $F=1/4$.  For a large superlattice consisting of many strips gated to the Dirac point interspersed among doped regions, we find $F=1/(8\ln{2})$.  These results differ from the value $F=1/3$ predicted for a disordered metal, providing an unambiguous experimental signature of evanescent mode transport in graphene.
\end{abstract}

\maketitle

\section{Introduction}

The nonequilibrium current fluctuations, or shot noise, in ballistic graphene \cite{novoselov:sci04,zhang:nat05,berger:jpc04,neto:rmp09} have received much attention since the seminal paper by Tworzydlo {\em et al.}~\cite{tworzydlo:prl06} In that work, it was shown that shot noise can be generated even in a completely impurity-free sheet of graphene.  This result is surprising at first sight, as the shot noise vanishes in conductors without electron scattering. \cite{khlus:jetp87,lesovik:jetp89,reznikov:prl95,kumar:prl96}
The unanticipated noise is caused by evanescent (exponentially damped) waves that  backscatter electrons, even in clean graphene. 

 It has been shown by Tworzydlo {\em et al.}~\cite{tworzydlo:prl06} that in a clean sheet of graphene at its Dirac point, that is, at zero chemical potential, the shot noise is not only nonzero, but, moreover, it has  universal characteristics. The shot noise normalized by the mean current and expressed in units of the electron charge $e$ -- this quantity  is commonly referred to as the Fano factor $F$  --  takes the   value $F=1/3$. This prediction has generated much theoretical interest in the topic, \cite{cayssol:prb09,rycerz:prb09,snyman:prb07,sanjose:prb07,lewenkopf:prb08,schuessler:prb09,sonin:prb08,golub:prb10} and considerable experimental activity.~\cite{dicarlo:prl08,danneau:prl08}

So far, the prediction of Tworzydlo {\em et al.}~\cite{tworzydlo:prl06} has been tested by two experiments, as reported in Refs.\ \onlinecite{dicarlo:prl08,danneau:prl08}, and a Fano factor close to $F=1/3$ was measured.  However, the interpretation of these experiments is ambiguous, as the universal shot noise value for evanescent mode transport  at the Dirac point of graphene\cite{tworzydlo:prl06} is identical to the one expected for conventional  disordered conductors in the diffusive regime.~\cite{beenakker:prl92} It has been confirmed in numerous regimes, both with \cite{golub:prb10} and without \cite{sanjose:prb07,lewenkopf:prb08,schuessler:prb09,sonin:prb08} electron-electron interactions, that a Fano factor close to $F=1/3$ is also expected for disordered graphene, independent of the chemical potential.  Therefore, the measured Fano factor $F=1/3$ could be due to either evanescent mode transport or impurities in the measured samples.  

In the experiment of Danneau {\em et al.},~\cite{danneau:prl08} an additional signature of evanescent mode transport was observed; namely a strong dependence of the measured Fano factor on the chemical potential.  Such dependence on the chemical potential is expected for clean graphene.~\cite{tworzydlo:prl06,schuessler:prb09,sonin:prb08}  As the chemical potential departs from the Dirac point, more and more evanescent waves become propagating and cease to backscatter electrons, decreasing the Fano factor. The observed dependence of $F$ on the chemical potential does not occur in generic diffusive conductors, \cite{sanjose:prb07,lewenkopf:prb08,schuessler:prb09,sonin:prb08} making it a more distinctive signature of evanescent mode transport. Nevertheless, other scenarios are also consistent with the doping dependence of $F$ reported by Danneau {\em et al.},~\cite{danneau:prl08} such as energy-dependent scattering. Additional experimental signatures of transport through evanescent modes in graphene are therefore desirable.

In this article, we propose experimental geometries for which the transport through evanescent modes at the Dirac point of graphene has unambiguous signatures. We first study a sheet of graphene subject to gate potentials that induce two strips of graphene with chemical potential at the Dirac point, separated by a highly doped region.  We show that the Fano factor takes a universal value in this geometry, as in that of Ref.\  \onlinecite{tworzydlo:prl06}: $F=1/4$. This value is different from that of diffusive conductors, and it provides an unambiguous signature of evanescent modes. Graphene devices with  multiple potential steps, as required for the proposed test of evanescent transport, have been implemented experimentally.~\cite{stander:prl09}

Similar results are obtained  for longer cascades of $p>2$ strips of graphene  with evanescent transport. In particular, we take the limit $p\to\infty$ of a long graphene superlattice with a piecewise constant potential. Such superlattices have recently received much attention as a way to engineer the bandstructure of graphene \cite{esmailpour:08,vanevic:prb09,park:nal08,brey:prl09,park:prl09,pereira:apl07,barbier:prb08,barbier:prb09,bliokh:prb09,arovas:10}  -  to the point of creating new Dirac cones in the electronic spectrum.  \cite{brey:prl09,park:prl09,pereira:apl07,barbier:prb08,barbier:prb09,bliokh:prb09,arovas:10}  Under certain conditions, we find for such superlattices yet another universal value of the Fano factor: $F=1/(8\ln{2})$. 

This paper is organized as follows: In section \ref{double}, we analytically derive the universal Fano factor $F=1/4$ for two evanescent regions in series. We study numerically the departures of $F$ from this  value as asymmetries are introduced into the setup, an additional experimental signature of   evanescent mode transport. In section \ref{superlattice}, we first generalize our approach to cascades of an arbitrary number $p$ of evanescent regions in series before taking the long superlattice limit $p\to\infty$. We conclude with a summary in section \ref{conclusions}.
\section{Evanescent mode transport through two evanescent strips in series} \label{double}
\begin{figure}
\centering
\includegraphics[width=\columnwidth]{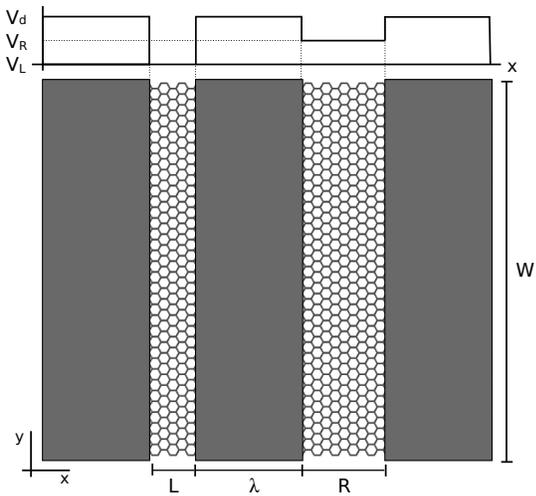}
\caption{Schematic of a graphene strip of width $W$ consisting of a highly doped region (gray rectangle) of length $\lambda$ at voltage $V_d$ separated by  weakly doped regions of length $L$ and $R$. Separate gate leads (not shown) control the voltages $V_L$ and $V_R$ in the weakly doped regions.  The system is contacted at either end by ideal leads (not shown) at the lead potential $V_\ell$.  The electrostatic potential in the strip is plotted above the schematic as a function of $x$.}
\label{fig:DoubleWellSchematic}
\end{figure}

We analyze the transport through a graphene sheet of width $W$ subject to gates that allow one to tune   two strips of lengths $L$ and $R$ to a chemical potential close to the Dirac point. We denote the electrostatic potentials in those strips by $V_L$ and $V_R$, respectively. The two strips are separated by a highly doped region of length $\lambda$ at electric potential  $V_d$  (see Fig.~\ref{fig:DoubleWellSchematic}).  

We model the leads that contact the sample by highly doped regions of graphene to the left and to the right of the sample at voltage $V_\ell$, as in the calculations of Tworzydlo {\em et al.}~\cite{tworzydlo:prl06} It has been confirmed in density functional calculations that this model correctly describes transport into certain types of contacts.~\cite{barraza:prl10}  In units where $\hbar =1$, our model Hamiltonian takes the form
\beq \label{Dirac}
H_\gamma= v \boldsymbol{\sigma_\gamma} \cdot  \boldsymbol{p} + V(x),
\eeq
where $\boldsymbol{ \sigma}_\gamma=(\gamma\sigma_x,\sigma_y)$ is a vector of Pauli matrices with the valley index $\gamma = \pm 1$, $\boldsymbol{p}$ is the electron momentum and  $v$ is the electron velocity.  The potential $V(x)$ takes the value $V_d$ in the highly doped regions, $V_\ell$ in the leads and the gate voltages $V_L$ and $V_R$ in the regions that can be tuned to the Dirac point (see Fig.~\ref{fig:DoubleWellSchematic}). We assume the microscopic potential, which is represented by the  potential $V(x)$ in our long-wavelength theory, to be smooth on the lattice scale, so that it does not scatter between valleys.

The plane wave solutions of the Dirac equation have energy $\epsilon = V_d\pm v\sqrt{k_d^2+q_n^2}$ in the highly doped regions and $\epsilon = V_{L(R)}\pm v\sqrt{k_{L(R)}^2+q_n^2}$ in the left (right) gated regions, where the $\pm$ sign refers to the conduction and valence bands, respectively.  We  choose $\left|V_{\ell (d)}\right|\gg v/L, v/R$ such that, at the Fermi level, all relevant modes are propagating (real $k_{\ell (d)}$) in the leads and the highly doped regions.  In the weakly doped regions, both propagating (real $k_{L(R)}$) and evanescent (imaginary $k_{L(R)}$) modes can occur.

The transverse wavenumbers $q_n$ depend on the boundary conditions at $y=0$ and $y=W$.  We consider a class of boundary conditions that do not couple longitudinal and transverse wavenumbers.~\cite{tworzydlo:prl06}  In this case, the transverse wavenumber $q_n$ is a good quantum number, and modes are not mixed at the weakly-to-highly doped region interfaces.  For the infinite mass, metallic armchair edge, or semiconducting armchair edge boundary conditions, the transverse wavenumber is given by $q_n = \left(n+\alpha\right)\pi/W$, where \cite{tworzydlo:prl06}
\beq
\alpha = \left\{ \begin{array}{ll}
1/2 & \mbox{infinite mass} \nn \\
0 & \mbox{metallic armchair} \nn \\
1/3 & \mbox{semiconducting armchair}. \end{array}\right. 
\eeq
In the continuum limit $W \gg L,R$, which we take for the remainder of this article, one expects  the transport properties to be independent of the boundary conditions. Accordingly, $\alpha$ drops out of the calculation. 
\subsection{The universal limit} \label{sub1}
We first consider the situation in which both weakly doped regions of the sample are tuned to the Dirac point, $V_L=V_R=0$. Matching modes at the  interfaces between segments of differing potential results in the transmission amplitude 
\beqa
t_n &= \left[ e^{ik_d\lambda}\sinh(n \mathcal{L})\sinh(n \mathcal{R})\right. + \nn \\
&\left. + e^{-ik_d\lambda}\cosh(n \mathcal{L})\cosh(n \mathcal{R})\right]^{-1} 
\label{eq:transmission_amplitude}
\eeqa
for mode $n$ at the Fermi level, which we choose to be at energy $\epsilon_{\rm F}=0$.  The dimensionless ``lengths''  $\mathcal{L}$ and $\mathcal{R}$ are given by $\mathcal{L}=\pi L/W$ and $\mathcal{R}=\pi R/W$.
The Fano factor is found from the transmission probabilities $T_n=|t_n|^2$ as \cite{lesovik:jetp89, buettiker:prl90}
\beq
F = \frac{\sum_n T_n(1-T_n)}{\sum_{n} T_n}.
\label{eq:Fano}
\eeq
  In the continuum limit, $W\gg L,R$, the sums become integrals over the mode index, and the Fano factor is given by $F = 1 - I_2/I_1$ with
\beqa
I_1 &= \int_{0}^{\infty} T_n dn, \nn \\
I_2 &= \int_{0}^{\infty} T_n^2 dn.
\label{eq:integrals}
\eeqa

\begin{figure}
\centering
\includegraphics[width=\columnwidth]{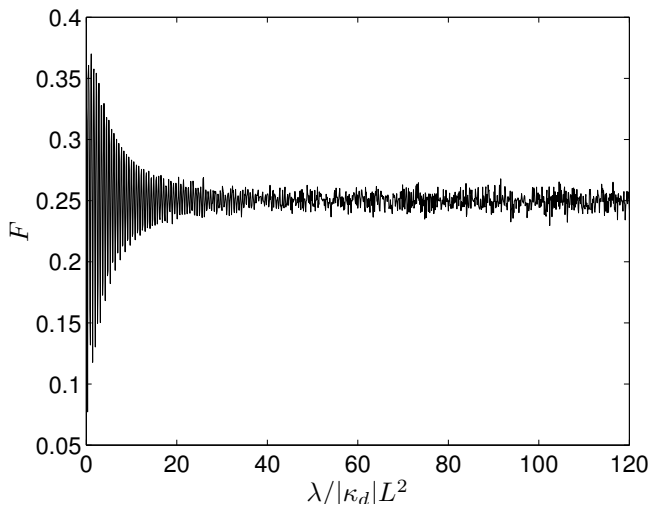}
\caption{Fano factor $F$ for two graphene strips as in Fig.~\ref{fig:DoubleWellSchematic} for $L=R$ and chemical potential at the Dirac point, as a function of the length of the highly doped region in units of \mbox{$\left|\kappa_d\right|L^2$}.  The curve is calculated from \Eq{eq:Fano} in the limit $N\rightarrow\infty$.  The Fano factor approaches $0.33\ldots$ as $\lambda\rightarrow 0$, in agreement with the calculations of Tworzydlo {\em et al.}~\cite{tworzydlo:prl06}  For large values of the length of the highly doped region, the Fano factor approaches $0.25\ldots$, in agreement with the analytic calculations presented in the text.}
\label{fig:Fano_of_lambda}
\end{figure}

In Fig.~\ref{fig:Fano_of_lambda}, the transmission probability determined from \Eq{eq:transmission_amplitude} is integrated numerically in order to obtain the Fano factor for a symmetric system ($L=R$)  as a function of the thickness  $\lambda$ of the central, highly doped region.  As $\lambda\rightarrow 0$, the Fano factor approaches $1/3$.  This result agrees with the calculations of Tworzydlo {\em et al.},~\cite{tworzydlo:prl06} as this limit corresponds to transmission through a single graphene strip  at the Dirac point.  In contrast, the Fano factor approaches $0.25\ldots$ in the limit $\lambda\gg\left|\kappa_d\right| L^2$, where $\kappa_d=V_d/v$.  

This numerical result suggests that the Fano factor may be accessible analytically in the limit $\lambda\gg\left|\kappa_d\right| L^2$. Indeed, one finds that the integrals in \Eq{eq:integrals} can be done analytically for $\lambda\gg\left|\kappa_d\right| L^2$.  Consider first $I_1$.  We write the transmission probability determined from \Eq{eq:transmission_amplitude}   in the form
\beq
T_n = \frac{1}{\alpha (n)}\frac{1}{1+\beta (n)\cos [2k_d(n)\lambda]},
\label{eq:T_of_lambda}
\eeq
where
\beqa
\alpha (n) &=& \Cosh{2}{n\mathcal{L}}\Cosh{2}{n\mathcal{R}} + \Sinh{2}{n\mathcal{L}}\Sinh{2}{n\mathcal{R}}, \nn \\
\beta (n) &=& 2\cosh (n\mathcal{L})\cosh (n\mathcal{R})\sinh (n\mathcal{L})\sinh (n\mathcal{R}) / \alpha(n), \nn \\
k_d(n) &=& \sqrt{\kappa_d^2-n^2\pi^2/W^2}.
\eeqa
The transmission probability decays exponentially with $n\left(\mathcal{L}+\mathcal{R}\right)$, and the integrals in \Eq{eq:integrals} are thus cut-off at $n\simeq \max\left\{ 1/\mathcal{L}, 1/\mathcal{R}\right\}$. 

The key observation is that for $\lambda\gg \left|\kappa_d\right| L^2$,  the cosine function in \Eq{eq:T_of_lambda}  oscillates rapidly on the scale of the exponential decay of $T_n$ (recall that we assume $\left|\kappa_d\right| L \gg 1 $). In the limit $\lambda/\left|\kappa_d\right|L^2\to \infty$, therefore, $\alpha (n)$ and $\beta (n)$ are constant within one period of oscillation
\beq
\Delta n(n_0) = \frac{\kappa_dW^2}{\lambda\pi n_0}
\label{eq:Delta_n}
\eeq
of the $\cos{\left[ 2k_d(n)\lambda\right]}$ function around a given index $n_0$. Consequently, $I_1$, restricted to an interval of length $\Delta  n(n_0)$ around $n_0$, becomes the integral of the function $\left[1+\beta (n)\cos{(\delta-\omega n)}\right]^{-1}$, which can be done analytically. Here, $\delta$ and $\omega$ are found by linearizing $2k_d(n)\lambda = \delta (n_0)-\omega (n_0) n$ with $\delta (n_0) = 2\kappa_d\lambda\left(1+n_0^2\pi^2/2\kappa_d^2W^2\right)$ and $\omega (n_0) = 2\pi^2\lambda n_0/\kappa_dW^2$.
 In the limit $\lambda/\left|\kappa_d\right|L^2\to \infty$, the periods $\Delta  n(n_0)$ are short compared to the decay scale of the transmission probability set by $\mathcal{L}+\mathcal{R}$, and the requisite sum over all periods becomes a second integral.  Noting that $|\beta (n_0)|\leq 1$, this results in
\beq
I_1 \approx\int_{0}^{\infty}\frac{1}{\gamma_- (n_0)} dn_0 ,
\label{eq:I1approx}
\eeq
where
\beqa
\gamma_\pm (n_0) &=& \Cosh{2}{n_0\mathcal{L}}\Cosh{2}{n_0\mathcal{R}}\pm \nn \\
&\pm &\Sinh{2}{n_0\mathcal{L}}\Sinh{2}{n_0\mathcal{R}}.
\eeqa
The integral $I_2$ of \Eq{eq:integrals} can be  done in the same manner, with the result
\beq
I_2\approx\int_{0}^{\infty}\frac{\gamma_+(n_0)}{\gamma_-^3(n_0)} dn_0.
\label{eq:I2approx}
\eeq
 
\begin{figure}
\centering
\includegraphics[width=\columnwidth]{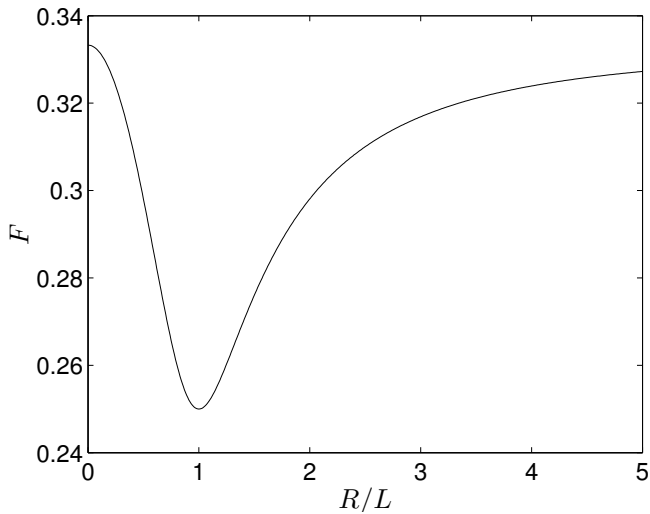}
\caption{Length ratio dependence of the Fano factor for two graphene strips as in Fig.~\ref{fig:DoubleWellSchematic} with chemical potential at the Dirac point.  The curve is calculated from the integrals given in \Eq{eq:I1approx} and \Eq{eq:I2approx} in the continuum limit with \mbox{$\lambda\gg \left|\kappa_d\right|L^2$}.  The Fano factor shows a minimum of $0.25\ldots$ for the symmetric case ($R/L=1$) and tends toward $0.33\ldots$ as $R/L\rightarrow 0$ or $R/L\rightarrow \infty$, in agreement with the results of Tworzydlo {\em et al.}~\cite{tworzydlo:prl06}}
\label{fig:Fano_of_lengths}
\end{figure}

In the symmetric case, $L=R$, the integrals in Eq.\  (\ref{eq:I1approx}) and (\ref{eq:I2approx}) can also be evaluated analytically, resulting in the same Fano factor as for a ballistic quantum dot, \cite{jalabert94} $F = 1/4$, in accordance with the numerical results shown in Fig.~\ref{fig:Fano_of_lambda}. Evanescent transport in this geometry thus has an unambiguous signature, with a Fano factor that differs from the one in a disordered sample. For the asymmetric case, we calculate the Fano factor numerically as a function of $R/L$, with the results plotted in Fig.~\ref{fig:Fano_of_lengths}.  The Fano factor approaches $1/3$ in the limit $R/L\rightarrow\infty$, which again corresponds to a single graphene region at the Dirac point, as considered by Tworzydlo {\em et al.}~\cite{tworzydlo:prl06} 

The dependence of the Fano factor on the ratio $R/L$ shown in Fig.~\ref{fig:Fano_of_lengths} is a more distinctive signature of evanescent wave transport than the value $F=1/4$ at $L=R$ alone.  In a typical experiment, however, the lengths of the gated regions cannot be changed easily.  Alternatively, the gate voltages $V_L$ and $V_R$ can be controlled. We discuss the dependence of $F$ on those gate voltages  in the following subsection.

\subsection{Voltage induced signatures of evanescent mode transport}

\begin{figure}
\centering
\includegraphics[width=\columnwidth]{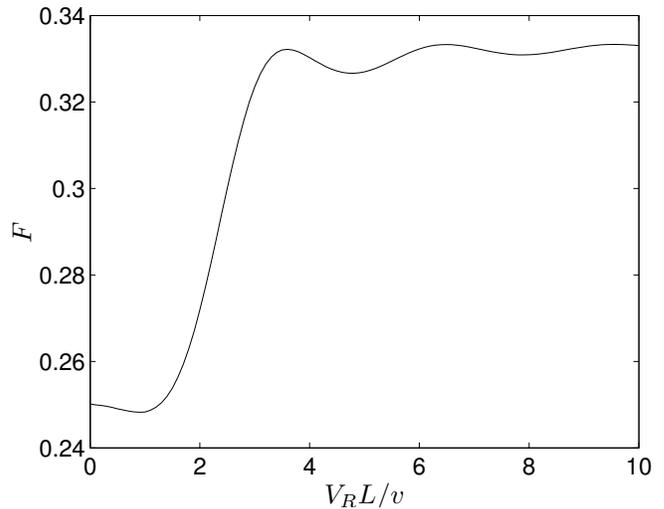}
\caption{The Fano factor for two graphene strips as in Fig.~\ref{fig:DoubleWellSchematic} for $L=R$ and  \mbox{$\lambda\gg \left|\kappa_d\right|L^2$} as a function of gate voltage $V_R$ in the right graphene region, measured in units of $v/L$.  Gate leads fix the chemical potential in the left evanescent region to the Dirac point ($V_L=0$).    The Fano factor is $0.25\ldots$ when $V_R$ is at the Dirac point, as previously calculated, and it approaches $0.33\ldots$ as $V_R\gg v/L$. }
\label{fig:Fano_of_V}
\end{figure}

Here, we obtain the dependence of the Fano factor on the gate voltage $V_R$ in the symmetric configuration, $L=R$, (cf.\  Fig.~\ref{fig:DoubleWellSchematic}) with the left gated region at the Dirac point, $V_L$=0.  The transmission amplitude as a function of $k_R$ takes the form
\beqa
t_n= k_R\left[ e^{-ik_d\lambda}\cosh{n\mathcal{L}}\left( k_R\cos{k_R L} - \right.\right. \nn \\
\left. -i\kappa_R\sin{k_R L}\right) + \left. e^{ik_d\lambda}\frac{n\pi}{W}\sinh{n \mathcal{L}}\sin{k_R L}\right]^{-1},
\label{eq:trans_of_V}
\eeqa
where $\kappa_{R(d)}^2=k_{R(d)}^2+q_n^2$ with $\kappa_{R(d)}=V_{R(d)}/v$.  We consider the limit $\lambda/\left|\kappa_d\right|L^2\to \infty$ of the resulting expression for the Fano factor, following the methods of the previous section.  In this case, we perform the final integration, corresponding to the sum over periods $\Delta n(n_0)$,  numerically, with the result plotted in Fig.~\ref{fig:Fano_of_V}.  The Fano factor is  $1/4$ when the gate voltage $V_R$ is at the Dirac point, as previously calculated, and it approaches $1/3$ in the limit $V_R\gg v/L$.  The crossover between $F=1/3$ and $F=1/4$ within a gate voltage interval of $\Delta V_L \simeq v/L$ serves as another distinctive signature of evanescent mode transport.  

\section{Evanescent mode transport in a graphene superlattice} \label{superlattice}
\begin{figure}
\centering
\includegraphics[width=\columnwidth]{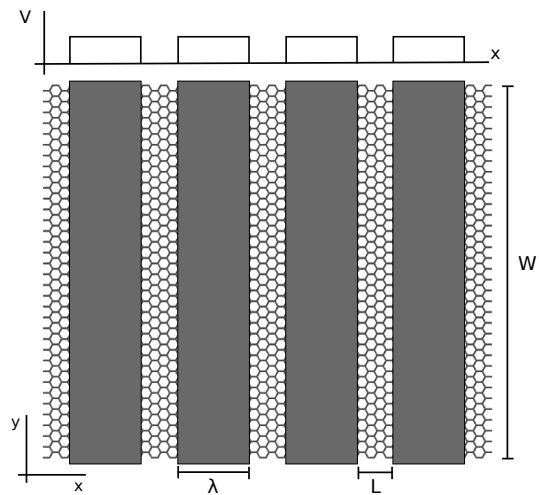}
\caption{Schematic of a segment of a graphene superlattice of width $W$, consisting of evanescent graphene regions of length $L$ separated by  doped regions (gray rectangles) of length $\lambda$ at the potential $V_d$.  Separate gate leads (not shown) fix the voltages in the graphene regions to the Dirac point, $V=0$.  The superlattice is contacted at either end by leads at potential $V_\ell$ (not shown).  The superlattice potential is plotted above the schematic as a function of $x$.}
\label{fig:RibbonSchematic}
\end{figure}

In this section, we investigate the shot noise for evanescent transport through a graphene superlattice consisting of many graphene strips tuned to the Dirac point, alternating with  doped regions, as depicted in Fig.~\ref{fig:RibbonSchematic}. We eventually take the long superlattice limit of an infinite number of such regions.   We take the graphene regions gated to the Dirac point to be of length $L$, and they are separated by doped regions of length $\lambda$ at the voltage $V_d$.  The contacts at both ends are again modeled by highly doped graphene at the lead potential $V_\ell$. The potential $V_\ell$ drops out of the calculation in the limit $V_\ell,V_d\gg v/L$.

\subsection{Transmission through a cascade of evanescent strips in series}
In order to calculate the transmission probability through the graphene superlattice depicted in Fig.~\ref{fig:RibbonSchematic}, we employ the transfer matrix method.~\cite{falko:prb06}  The transfer matrix $M(x,x')$ for the requisite Dirac spinors of a mode with transverse momentum $q$ satisfies the equation
\beq
\partial_x M(x,x') = \left[ -i\frac{V(x)}{v}\sigma_x+q\sigma_z\right] M(x,x').
\label{eq:DiracTransfer}
\eeq
In addition, one has~\cite{falko:prb06} $M(x,x)=I$, $M(x,x')=M(x,x'')M(x'',x')$, $\det M(x,x')=1$ and $M^\dagger (x,x')\sigma_xM(x,x')=\sigma_x$. The latter condition ensures current conservation.

We write the transfer matrix in the  doped regions as $M_d(x,x')=A_d(x)A_d^{-1}(x')$. Here, $A_d(x)$ satisfies \Eq{eq:DiracTransfer} in the doped regions, where $V(x)=V_d$.  Similarly, the transfer matrix in the lead regions contacting either end of the superlattice is $M_\ell(x,x')=A^\pdag_\ell(x)A_\ell^{-1}(x')$, where $A_\ell(x)$ satisfies \Eq{eq:DiracTransfer} in the lead regions with $V(x)=V_\ell $. Following Ref.\ \onlinecite{falko:prb06}, we choose matrices $A_{\ell (d)}(x)$ that allow one to project onto right and left moving states:
\beq
A_j(x) = \sqrt{\frac{\kappa_j}{2k_j}}\left(\begin{array}{cc} \frac{k_j+i q}{\kappa_j}e^{-ik_jx} & \frac{-k_j+i q}{\kappa_j}e^{ik_jx} \\  e^{-ik_jx} & e^{ik_jx}\end{array}\right),
\label{eq:Amatrix}
\eeq
where $j=\{ \ell ,d\}$ and  $k_j^2=\kappa_j^2-q^2$ with $\kappa_j=V_j/v$.

The transfer matrix in the evanescent regions, where the electric potential vanishes, is   $M_e(x,x')=\exp{\left[ q\sigma_z\left(x-x'\right)\right]}$.  The transfer matrix through a single sequence consisting of a region at the Dirac point followed by a  doped segment is then given by \mbox{$M_1(L+\lambda,0)=M_e(L+\lambda,\lambda)M_d(\lambda,0)$}.
The transfer matrix for $p$ doped-evanescent segments in series, contacted at either end by ideal leads, is found using $M_1$ and the matrices $A_{\ell }$ of \Eq{eq:Amatrix}  as
\beqa \label{Mp}
M_p(x,x')&=&A_\ell(x)A_\ell^{-1}[p(L+\lambda)]\left[M_1(L+\lambda,0)\right]^p\times \nn \\
&\times &A_\ell(0)A_\ell^{-1}(x')
\eeqa
at $x>p(L+\lambda)$ and $x'<0$. 

From the asymptotic form of $M_p(x,x')$ in the leads, one extracts the transmission amplitudes through the entire array as explained in Ref.\ \onlinecite{falko:prb06}: $T^{(p)}=1/\left|\alpha\right|^2$, where
\beq
\left(\begin{array}{cc}\alpha & \beta^\ast \\ \beta & \alpha^\ast\end{array}\right)=\lim_{x\to\infty}  A_\ell^{-1}(x ) M_p(x,-x) A_\ell(-x),
\label{eq:probmatrix}
\eeq
which, with Eq.\ (\ref{Mp}), becomes
\beq
\left(\begin{array}{cc}\alpha & \beta^\ast \\ \beta & \alpha^\ast\end{array}\right)= A_\ell^{-1}(p(L+\lambda))\left[M_1(L+\lambda,0)\right]^pA_\ell (0).
\eeq

The transmission probability $T_n^{(p)}$ of mode $n$ is obtained by setting $q=q_n$ in the above equations, and we find
\begin{widetext}
\beq
T_n^{(p)}= \frac{2\Delta^2(n)} {\left[ \Cosh{2}{n\mathcal{L}}-1\right] \cosh{\left[2p\cdot\arcsinh{\Delta(n)}\right]} + \left[ \cos{\left(2k_d(n)\lambda\right) }\Cosh{2}{n\mathcal{L}}-1 \right] }.
\label{eq:T_of_p}
\eeq
\end{widetext}
Here, $\Delta (n) = \sqrt{\cos^2{\left[k_d(n)\lambda\right]}\Cosh{2}{n\mathcal{L}}-1}$. As previously mentioned, the transmission probability   becomes independent of the lead potential in the limit $\kappa_\ell , \kappa_d\gg 1/L$ that we have taken. 
In the special case of $p=2$, this equation reduces to \Eq{eq:T_of_lambda}.

\begin{figure}
\centering
\includegraphics[width=\columnwidth]{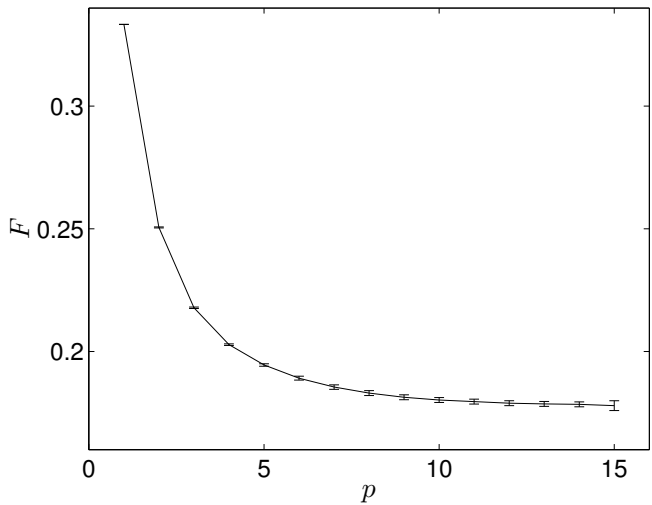}
\caption{Fano factor for a graphene superlattice, as pictured in Fig.~\ref{fig:RibbonSchematic}, as a function of $p$, the length of the cascade.  The curve is calculated from the transmission probability given in \Eq{eq:T_of_p} with $\lambda = 10| \kappa_d| L^2$.  The Fano factor starts at its maximum value of $1/3$ for a single graphene strip, as considered by Tworzydlo {\em et al.}~\cite{tworzydlo:prl06}  In the long superlattice limit, $p\rightarrow\infty$, the Fano factor approaches $0.180...$, in agreement with our analytic calculations. The error bars are caused by errors in the numerical integration.}
\label{fig:Fano_of_p}
\end{figure}

The Fano factor can be calculated numerically for a cascade of arbitrary length by substitution of the transmission probability (\ref{eq:T_of_p}) into the integrals $I_1$ and $I_2$ of \Eq{eq:integrals}.  The results are plotted as a function of the cascade size $p$ in Fig.~\ref{fig:Fano_of_p} with error bars caused by errors in the numerical integration.  The Fano factor decreases rapidly from its maximum value of $1/3$ for $p=1$ and approaches $0.180...$ as $p\rightarrow\infty$.  We next compute the exact value of $F$ in this limit $p\rightarrow\infty$ of a long superlattice, again assuming $\lambda\gg\left|\kappa_d\right| L^2$. 

\subsection{The long superlattice limit}
\begin{figure}
\centering
\includegraphics[width=\columnwidth]{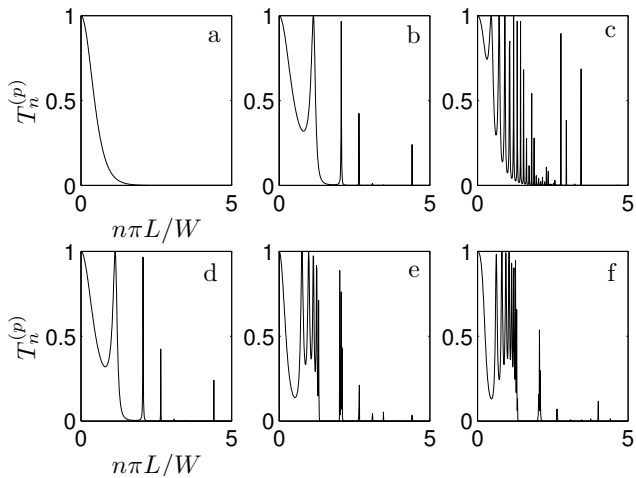}
\caption{Plots of the transmission probability \Eq{eq:T_of_p} as a function of mode index $n$.  The cascade size is  $p=2$ in the first row of plots, while the  doped region thickness takes the increasing values $\lambda = 0.01\left|\kappa_d\right| L^2$ (a),  $\lambda = 2\left|\kappa_d\right| L^2$ (b) and $\lambda=20\left|\kappa_d\right| L^2$ (c).  Larger thicknesses lead to a higher frequency of peaks in the transmission probability.  In the second row, the doped region thickness is  $\lambda=2\left|\kappa_d\right| L^2$, and the increasing cascade sizes $p=2$ (a), $p=6$ (b), and $p=10$ (c) cause an increasing frequency of the sub-peak oscillations and strong damping outside of the peak regions.}
\label{fig:transmission}
\end{figure}

The transmission probability $T_n^{(p)}$ of \Eq{eq:T_of_p} is plotted as a function of mode index $n$  for various values of $\lambda$ and $p$ in Fig.~\ref{fig:transmission}. One observes a series of peaks in  $T_n^{(p)}$, whose number increases with $\lambda /\left|\kappa_d\right| L^2$, as illustrated in the first row of plots in Fig.~\ref{fig:transmission}. The peaks occur at values of $n$ where $\Delta$ is imaginary, and $T_n^{(p)}$ is exponentially damped whenever $\Delta$ is a real number.  

When $\Delta$ is imaginary, the second hyperbolic cosine  in the denominator of $T_n^{(p)}$  oscillates with a frequency proportional to $p$, the number of graphene segments in the cascade.  Therefore, $T_n^{(p)}$ has a series of ``subpeaks'' in the regions of imaginary $\Delta$, whose number increases with $p$, as illustrated in the second row of plots in Fig.~\ref{fig:transmission}. 
We further observe in the second row of Fig.~\ref{fig:transmission} that  the damping to the sides of the regions with imaginary $\Delta$ is enhanced as $p$ increases. This is also due to the factor of $p$ in the second hyperbolic cosine function of \Eq{eq:T_of_p}.

\begin{figure}
\centering
\includegraphics[width=\columnwidth]{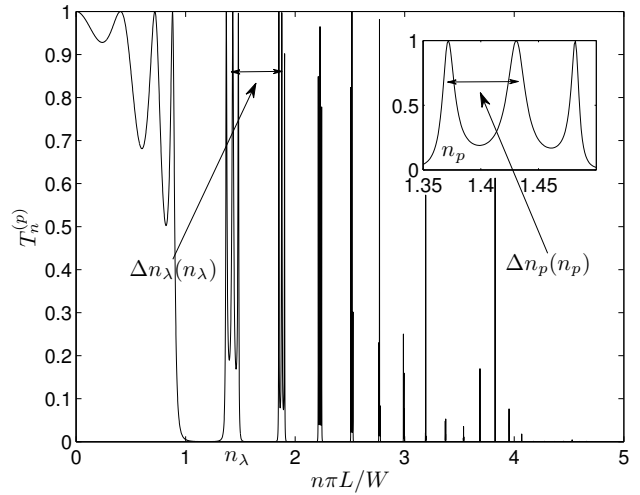}
\caption{Transmission probability \Eq{eq:T_of_p} as a function of mode index $n$  for $p=4$ and $\lambda = 4\left|\kappa_d\right| L^2$, showing peak oscillations with period set by $\lambda$ and sub-peak oscillations with period set by $p\lambda$.  The first peak region to the right of the central region, indexed by $n_\lambda$, is shown magnified in the inset.  The peak period is given by $\Delta n_\lambda (n_\lambda )$ in the neighborhood of peak $n_\lambda$, while the period of sub-peak oscillation is $\Delta n_p (n_p)$, around peaks indexed by $n_p$.  Both periods become infinitesimal in the limit $p\gg 1$ with $\lambda\gg\left|\kappa_d\right| L^2$.}
\label{fig:trans_inset}
\end{figure}

Motivated by the above observations, we partition the wave numbers into a series of peak regions where $\Delta$ is imaginary. The length of these regions is on the order of $\Delta n (n_\lambda )$ in \Eq{eq:Delta_n}.  Each peak region is then sub-partitioned into sub-peak regions, with a length that corresponds to the oscillation period of the second hyperbolic cosine of  \Eq{eq:T_of_p} at imaginary $\Delta$. That  length is of the order  \mbox{$\Delta n_p (n_p)\sim\Delta n (n_p )/p$}.   Fig.~\ref{fig:trans_inset} illustrates both periods of oscillation, showing the transmission probability \Eq{eq:T_of_p}  for $p=4$ with doped region thickness $\lambda = 4\left|\kappa_d\right| L^2$.

In order to calculate the Fano factor for a long graphene superlattice ($p\gg 1$), we first integrate $I_1$ and $I_2$ of \Eq{eq:integrals} for each  peak region $\Delta n_\lambda (n_\lambda )$ over all sub-peak regions $\Delta n_p(n_p)$.  The result is then summed over the peak region $\Delta n_\lambda (n_\lambda )$ by a second integration, resulting  in the desired integral over one peak region.  Finally, in the limit $\lambda\gg\left|\kappa_d\right| L^2$, the peak spacing becomes small compared to the decay scale of the transmission probability, set by $\mathcal{L}$, and the full integral can be found by further summing the single peak result over all such peaks, amounting to a third integration.

All of the above integration steps can be performed analytically, giving the exact   Fano factor  in the limit  $p\rightarrow\infty$ with $\lambda /\left|\kappa_d\right| L^2\rightarrow\infty$: $F=1/(8\ln{2})$.  This result is in agreement with the numerical results plotted in Fig.~\ref{fig:Fano_of_p}, and it differs from the Fano factor calculated for the geometry of section \ref{double}, as well as  the calculations of Tworzydlo \emph{et al.}~\cite{tworzydlo:prl06}  The shot noise for transport through a graphene superlattice as depicted in Fig.~\ref{fig:RibbonSchematic} thus provides another unambiguous signature of evanescent mode transport.  

\section{Discussion and Conclusions} \label{conclusions}
Throughout this article, we have assumed translational invariance in the y-direction. We now briefly discuss the sensitivity of our results to a breaking of that symmetry before summarizing. As a typical mechanism for such symmetry breaking, we consider a gate edge which makes  a non-zero angle $\Delta\phi$ with  the vertical. At such an interface, transverse modes are mixed. Electrons with momentum $k_x\sim\kappa_d=V_d /v$ in the $x$-direction increase their momentum in the $y$-direction by an amount $\Delta q  \approx \kappa_d\Delta\phi$ as they traverse such an edge.

To avoid qualitatively changing the above calculations, this momentum shift must be small on the scale $ \kappa_d/p\lambda q$ of the oscillations of the transmission probability, requiring $ \Delta\phi\ll 1/p\lambda q$. Noting that the typical  momenta contributing to the Fano factor in our calculation are of order $q \simeq 1/L$, this results in the condition
\beq
\Delta\phi\ll  L/p\lambda.
\eeq
In addition, we have assumed  $\lambda\gg\left|\kappa_d\right|L^2$ in all of our analytical calculations. These results therefore require 
\beq
1\ll \left|\kappa_d\right|L\ll\frac{\lambda}{L}\ll\frac{1}{p\Delta\phi}
\eeq
in setups with angular imperfections. 

In conclusion, we have proposed several experimental geometries for which the shot noise provides an unambiguous signature of transport through evanescent modes at the Dirac point of graphene.  For the case of two gated  graphene strips in series, separated by a long highly doped region, the Fano factor can be controlled through either spatial asymmetry or the gate potentials.  In particular, the Fano factor has a universal  minimum of $F=1/4$ for the spatially symmetric case with both gated regions tuned to the Dirac point.  This result differs from the value $F=1/3$ expected for transport through a disordered metal, which allows for a particularly conclusive experimental investigation of evanescent mode transport in graphene.  For the case of a long superlattice of evanescent regions we predict another universal value of the Fano factor, $F=1/(8\ln{2})$.

\acknowledgements

We gratefully acknowledge discussions with W.\ de Heer and B.\ Trauzettel. This work was funded by the NSF (DMR-0820382). 



\begin{thebibliography}{19}
\expandafter\ifx\csname natexlab\endcsname\relax\def\natexlab#1{#1}\fi
\expandafter\ifx\csname bibnamefont\endcsname\relax
  \def\bibnamefont#1{#1}\fi
\expandafter\ifx\csname bibfnamefont\endcsname\relax
  \def\bibfnamefont#1{#1}\fi
\expandafter\ifx\csname citenamefont\endcsname\relax
  \def\citenamefont#1{#1}\fi
\expandafter\ifx\csname url\endcsname\relax
  \def\url#1{\texttt{#1}}\fi
\expandafter\ifx\csname urlprefix\endcsname\relax\def\urlprefix{URL }\fi
\providecommand{\bibinfo}[2]{#2}
\providecommand{\eprint}[2][]{\url{#2}}

\bibitem[{\citenamefont{Novoselov et~al.}(2004)\citenamefont{Novoselov, Geim,
  Morozov, Jiang, Zhang, Dubonos, Grigorieva, and Firsov}}]{novoselov:sci04}
\bibinfo{author}{\bibfnamefont{K.S.}~\bibnamefont{Novoselov}},
  \bibinfo{author}{\bibfnamefont{A.~K.}~\bibnamefont{Geim}},
  \bibinfo{author}{\bibfnamefont{S.~V.}~\bibnamefont{Morozov}},
  \bibinfo{author}{\bibfnamefont{D.}~\bibnamefont{Jiang}},
  \bibinfo{author}{\bibfnamefont{Y.}~\bibnamefont{Zhang}},
  \bibinfo{author}{\bibfnamefont{S.~V.}~\bibnamefont{Dubonos}},
  \bibinfo{author}{\bibfnamefont{I.~V.}~\bibnamefont{Grigorieva}},
  \bibnamefont{and} \bibinfo{author}{\bibfnamefont{A.~A.}~\bibnamefont{Firsov}},
  \bibinfo{journal}{Science} \textbf{\bibinfo{volume}{306}},
  \bibinfo{pages}{666} (\bibinfo{year}{2004}).

\bibitem[{\citenamefont{Zhang et~al.}(2005)\citenamefont{Zhang, Tan, Stormer,
  and Kim}}]{zhang:nat05}
\bibinfo{author}{\bibfnamefont{Y.}~\bibnamefont{Zhang}},
  \bibinfo{author}{\bibfnamefont{Y.-W.} \bibnamefont{Tan}},
  \bibinfo{author}{\bibfnamefont{H.~L.} \bibnamefont{Stormer}},
  \bibnamefont{and} \bibinfo{author}{\bibfnamefont{P.}~\bibnamefont{Kim}},
  \bibinfo{journal}{Nature} \textbf{\bibinfo{volume}{438}},
  \bibinfo{pages}{201} (\bibinfo{year}{2005}).

\bibitem[{\citenamefont{Berger et~al.}(2004)\citenamefont{Berger, Song, Li, Li,
  Ogbazghi, Feng, Dai, Marchenkov, Conrad, First et~al.}}]{berger:jpc04}
\bibinfo{author}{\bibfnamefont{C.}~\bibnamefont{Berger}},
  \bibinfo{author}{\bibfnamefont{Z.}~\bibnamefont{Song}},
  \bibinfo{author}{\bibfnamefont{T.}~\bibnamefont{Li}},
  \bibinfo{author}{\bibfnamefont{X.}~\bibnamefont{Li}},
  \bibinfo{author}{\bibfnamefont{A.~Y.} \bibnamefont{Ogbazghi}},
  \bibinfo{author}{\bibfnamefont{R.}~\bibnamefont{Feng}},
  \bibinfo{author}{\bibfnamefont{Z.}~\bibnamefont{Dai}},
  \bibinfo{author}{\bibfnamefont{A.~N.} \bibnamefont{Marchenkov}},
  \bibinfo{author}{\bibfnamefont{E.~H.} \bibnamefont{Conrad}},
  \bibinfo{author}{\bibfnamefont{P.~N.} \bibnamefont{First}}, 
  \bibinfo{author}{\bibfnamefont{W.~A.} \bibnamefont{de Heer}},
  \bibinfo{journal}{J. Phys. Chem. B}
  \textbf{\bibinfo{volume}{108}}, \bibinfo{pages}{19912}
  (\bibinfo{year}{2004}).

\bibitem[{\citenamefont{Neto et~al.}(2009)\citenamefont{Neto, Guinea, Peres,
  Novoselov, and Geim}}]{neto:rmp09}
\bibinfo{author}{\bibfnamefont{A.~H.} \bibnamefont{Castro Neto}},
  \bibinfo{author}{\bibfnamefont{F.}~\bibnamefont{Guinea}},
  \bibinfo{author}{\bibfnamefont{N.~M.~R.} \bibnamefont{Peres}},
  \bibinfo{author}{\bibfnamefont{K.~S.} \bibnamefont{Novoselov}},
  \bibnamefont{and} \bibinfo{author}{\bibfnamefont{A.~K.} \bibnamefont{Geim}},
  \bibinfo{journal}{Rev. Mod. Phys.} \textbf{\bibinfo{volume}{81}},
  \bibinfo{eid}{109} (\bibinfo{year}{2009}).

\bibitem[{\citenamefont{Tworzydlo et~al.}(2006)\citenamefont{Tworzydlo,
  Trauzettel, Titov, Rycerz, and Beenakker}}]{tworzydlo:prl06}
\bibinfo{author}{\bibfnamefont{J.}~\bibnamefont{Tworzydlo}},
  \bibinfo{author}{\bibfnamefont{B.}~\bibnamefont{Trauzettel}},
  \bibinfo{author}{\bibfnamefont{M.}~\bibnamefont{Titov}},
  \bibinfo{author}{\bibfnamefont{A.}~\bibnamefont{Rycerz}}, \bibnamefont{and}
  \bibinfo{author}{\bibfnamefont{C.~W.~J.} \bibnamefont{Beenakker}},
  \bibinfo{journal}{Phys. Rev. Lett.} \textbf{\bibinfo{volume}{96}},
  \bibinfo{eid}{246802} (\bibinfo{year}{2006}).

\bibitem[{\citenamefont{Khlus}(1987)}]{khlus:jetp87}
\bibinfo{author}{\bibfnamefont{V.~A.}~\bibnamefont{Khlus}},
  \bibinfo{journal}{Zh.  \'Eksp. Teor. Fiz.} \textbf{\bibinfo{volume}{ 93}},
  \bibinfo{pages}{2179} (\bibinfo{year}{1987}).
[\bibinfo{journal}{Sov. Phys. JETP} \textbf{\bibinfo{volume}{66}},
  \bibinfo{pages}{1243} (\bibinfo{year}{1987})].

\bibitem[{\citenamefont{Lesovik}(1989)}]{lesovik:jetp89}
\bibinfo{author}{\bibfnamefont{G.~B.}~\bibnamefont{Lesovik}},
\bibinfo{journal}{Pis'ma Zh. \'Eksp. Teor. Fiz.} \textbf{\bibinfo{volume}{49}},
  \bibinfo{pages}{513} (\bibinfo{year}{1989})
  [\bibinfo{journal}{JETP Lett.} \textbf{\bibinfo{volume}{49}},
  \bibinfo{pages}{592} (\bibinfo{year}{1989})].

\bibitem[{\citenamefont{Reznikov et~al.}(1995)\citenamefont{Reznikov, Heiblum,
  Shtrikman, and Mahalu}}]{reznikov:prl95}
\bibinfo{author}{\bibfnamefont{M.}~\bibnamefont{Reznikov}},
  \bibinfo{author}{\bibfnamefont{M.}~\bibnamefont{Heiblum}},
  \bibinfo{author}{\bibfnamefont{H.}~\bibnamefont{Shtrikman}},
  \bibnamefont{and} \bibinfo{author}{\bibfnamefont{D.}~\bibnamefont{Mahalu}},
  \bibinfo{journal}{Phys. Rev. Lett.} \textbf{\bibinfo{volume}{75}},
  \bibinfo{pages}{3340} (\bibinfo{year}{1995}).

\bibitem[{\citenamefont{Kumar et~al.}(1995)\citenamefont{Kumar, Saminadayar,
  Glatti, Jin, and Etienne}}]{kumar:prl96}
\bibinfo{author}{\bibfnamefont{A.}~\bibnamefont{Kumar}},
  \bibinfo{author}{\bibfnamefont{L.}~\bibnamefont{Saminadayar}},
  \bibinfo{author}{\bibfnamefont{D.~C.}~\bibnamefont{Glattli}}, 
  \bibinfo{author}{\bibfnamefont{Y.}~\bibnamefont{Jin}},
  \bibnamefont{and} \bibinfo{author}{\bibfnamefont{B.}~\bibnamefont{Etienne}},
  \bibinfo{journal}{Phys. Rev. Lett.} \textbf{\bibinfo{volume}{76}},
  \bibinfo{pages}{2778} (\bibinfo{year}{1996}).

\bibitem[{\citenamefont{Cayssol et~al.}(2009)\citenamefont{Cayssol, Huard,
  and Goldhaber-Gordon}}]{cayssol:prb09}
\bibinfo{author}{\bibfnamefont{J.}~\bibnamefont{Cayssol}},
  \bibinfo{author}{\bibfnamefont{B.} \bibnamefont{Huard}},
  \bibnamefont{and} \bibinfo{author}{\bibfnamefont{D.}
  \bibnamefont{Goldhaber-Gordon}}, \bibinfo{journal}{Phys. Rev. B} \textbf{\bibinfo{volume}{79}},
  \bibinfo{pages}{075428} (\bibinfo{year}{2009}).

\bibitem[{\citenamefont{Rycerz et~al.}(2009)\citenamefont{Rycerz,
  Recher, and Wimmer}}]{rycerz:prb09}
\bibinfo{author}{\bibfnamefont{A.}~\bibnamefont{Rycerz}},
  \bibinfo{author}{\bibfnamefont{P.}~\bibnamefont{Recher}},
  \bibnamefont{and} \bibinfo{author}{\bibfnamefont{M.} \bibnamefont{Wimmer}},
  \bibinfo{journal}{Phys. Rev. B} \textbf{\bibinfo{volume}{80}},
  \bibinfo{eid}{125417} (\bibinfo{year}{2009}).

\bibitem[{\citenamefont{Snyman et~al.}(2007)\citenamefont{Snyhman,
  and Beenakker}}]{snyman:prb07}
\bibinfo{author}{\bibfnamefont{I.}~\bibnamefont{Snyman}},
  \bibnamefont{and} \bibinfo{author}{\bibfnamefont{C.~W.~J.} \bibnamefont{Beenakker}},
  \bibinfo{journal}{Phys. Rev. B} \textbf{\bibinfo{volume}{75}},
  \bibinfo{eid}{045322} (\bibinfo{year}{2007}).

\bibitem[{\citenamefont{San-Jose et~al.}(2007)\citenamefont{San-Jose, Prada,
  and Golubev}}]{sanjose:prb07}
\bibinfo{author}{\bibfnamefont{P.}~\bibnamefont{San-Jose}},
  \bibinfo{author}{\bibfnamefont{E.}~\bibnamefont{Prada}}, \bibnamefont{and}
  \bibinfo{author}{\bibfnamefont{D.~S.} \bibnamefont{Golubev}},
  \bibinfo{journal}{Phys. Rev. B} \textbf{\bibinfo{volume}{76}},
  \bibinfo{pages}{195445} (\bibinfo{year}{2007}).

\bibitem[{\citenamefont{Lewenkopf et~al.}(2008)\citenamefont{Lewenkopf,
  Mucciolo, and Castro~Neto}}]{lewenkopf:prb08}
\bibinfo{author}{\bibfnamefont{C.~H.} \bibnamefont{Lewenkopf}},
  \bibinfo{author}{\bibfnamefont{E.~R.} \bibnamefont{Mucciolo}},
  \bibnamefont{and} \bibinfo{author}{\bibfnamefont{A.~H.}
  \bibnamefont{Castro~Neto}}, \bibinfo{journal}{Phys. Rev. B}
  \textbf{\bibinfo{volume}{77}}, \bibinfo{pages}{081410(R)}
  (\bibinfo{year}{2008}).
  
\bibitem[{\citenamefont{Schuessler et~al.}(2009)\citenamefont{Schuessler,
  Ostrovsky, Gornyi, and Mirlin}}]{schuessler:prb09}
\bibinfo{author}{\bibfnamefont{A.}~\bibnamefont{Schuessler}},
  \bibinfo{author}{\bibfnamefont{P.~M.} \bibnamefont{Ostrovsky}},
  \bibinfo{author}{\bibfnamefont{I.~V.} \bibnamefont{Gornyi}},
  \bibnamefont{and} \bibinfo{author}{\bibfnamefont{A.~D.}
  \bibnamefont{Mirlin}}, \bibinfo{journal}{Phys. Rev. B}
  \textbf{\bibinfo{volume}{79}}, \bibinfo{pages}{075405}
  (\bibinfo{year}{2009}).

\bibitem[{\citenamefont{Sonin}(2008)}]{sonin:prb08}
\bibinfo{author}{\bibfnamefont{E.~B.} \bibnamefont{Sonin}},
  \bibinfo{journal}{Phys. Rev. B} \textbf{\bibinfo{volume}{77}},
  \bibinfo{pages}{233408} (\bibinfo{year}{2008}).

\bibitem[{\citenamefont{Golub and Horovitz}(2010)}]{golub:prb10}
\bibinfo{author}{\bibfnamefont{A.}~\bibnamefont{Golub}} \bibnamefont{and}
  \bibinfo{author}{\bibfnamefont{B.}~\bibnamefont{Horovitz}},
  \bibinfo{journal}{Phys. Rev. B} \textbf{\bibinfo{volume}{81}},
  \bibinfo{pages}{245424} (\bibinfo{year}{2010}).

\bibitem[{\citenamefont{DiCarlo et~al.}(2008)\citenamefont{DiCarlo, Williams,
  Zhang, McClure, and Marcus}}]{dicarlo:prl08}
\bibinfo{author}{\bibfnamefont{L.}~\bibnamefont{DiCarlo}},
  \bibinfo{author}{\bibfnamefont{J.~R.} \bibnamefont{Williams}},
  \bibinfo{author}{\bibfnamefont{Y.}~\bibnamefont{Zhang}},
  \bibinfo{author}{\bibfnamefont{D.~T.} \bibnamefont{McClure}},
  \bibnamefont{and} \bibinfo{author}{\bibfnamefont{C.~M.}
  \bibnamefont{Marcus}}, \bibinfo{journal}{Phys. Rev. Lett.}
  \textbf{\bibinfo{volume}{100}}, \bibinfo{pages}{156801} (\bibinfo{year}{2008}).

\bibitem[{\citenamefont{Danneau et~al.}(2008)\citenamefont{Danneau, Wu,
  Craciun, Russo, Tomi, Salmilehto, Morpurgo, and Hakonen}}]{danneau:prl08}
\bibinfo{author}{\bibfnamefont{R.}~\bibnamefont{Danneau}},
  \bibinfo{author}{\bibfnamefont{F.}~\bibnamefont{Wu}},
  \bibinfo{author}{\bibfnamefont{M.~F.} \bibnamefont{Craciun}},
  \bibinfo{author}{\bibfnamefont{S.}~\bibnamefont{Russo}},
  \bibinfo{author}{\bibfnamefont{M.~Y.} \bibnamefont{Tomi}},
  \bibinfo{author}{\bibfnamefont{J.}~\bibnamefont{Salmilehto}},
  \bibinfo{author}{\bibfnamefont{A.~F.} \bibnamefont{Morpurgo}},
  \bibnamefont{and} \bibinfo{author}{\bibfnamefont{P.~J.}
  \bibnamefont{Hakonen}}, \bibinfo{journal}{Phys. Rev. Lett.}
  \textbf{\bibinfo{volume}{100}}, \bibinfo{pages}{196802}
  (\bibinfo{year}{2008}).
  
\bibitem[{\citenamefont{Beenakker and B\"uttiker}(1992)}]{beenakker:prl92}
\bibinfo{author}{\bibfnamefont{C.~W.~J.} \bibnamefont{Beenakker}}
  \bibnamefont{and}
  \bibinfo{author}{\bibfnamefont{M.}~\bibnamefont{B\"uttiker}},
  \bibinfo{journal}{Phys. Rev. B} \textbf{\bibinfo{volume}{46}},
  \bibinfo{pages}{1889} (\bibinfo{year}{1992}).  

\bibitem[{\citenamefont{Stander et~al.}(2009)\citenamefont{Stander, Huard,
 and Goldhaber-Gordon}}]{stander:prl09}
\bibinfo{author}{\bibfnamefont{N.}~\bibnamefont{Stander}},
  \bibinfo{author}{\bibfnamefont{B.}~\bibnamefont{Huard}},
  \bibnamefont{and} \bibinfo{author}{\bibfnamefont{D.}
  \bibnamefont{Goldhaber-Gordon}}, \bibinfo{journal}{Phys. Rev. Lett.}
  \textbf{\bibinfo{volume}{102}}, \bibinfo{pages}{026807}
  (\bibinfo{year}{2009}).

\bibitem[{\citenamefont{Esmailpour et~al.}(2008)\citenamefont{Esmailpour, Abedpour,
  Asgari, and Tabar}}]{esmailpour:08}
\bibinfo{author}{\bibfnamefont{N.}~\bibnamefont{Abedpour}},
  \bibinfo{author}{\bibfnamefont{A.}~\bibnamefont{Esmailpour}},
  \bibinfo{author}{\bibfnamefont{R.}~\bibnamefont{Asgari}}, \bibnamefont{and}
  \bibinfo{author}{\bibfnamefont{M.~R.} \bibnamefont{Tabar}},
  \bibinfo{journal}{Phys. Rev. B} \textbf{\bibinfo{volume}{79}},
  \bibinfo{pages}{165412} (\bibinfo{year}{2009}).

\bibitem[{\citenamefont{Vanevi\'{c} et~al.}(2009)\citenamefont{Vanevi\'{c},
  Stojanovi\'{c}, and Kindermann}}]{vanevic:prb09}
\bibinfo{author}{\bibfnamefont{M.}~\bibnamefont{Vanevi\'{c}}},
  \bibinfo{author}{\bibfnamefont{V.~M.} \bibnamefont{Stojanovi\'{c}}},
  \bibnamefont{and}
  \bibinfo{author}{\bibfnamefont{M.}~\bibnamefont{Kindermann}},
  \bibinfo{journal}{Phys. Rev. B} \textbf{\bibinfo{volume}{80}},
  \bibinfo{eid}{045410} (pages~\bibinfo{numpages}{8}) (\bibinfo{year}{2009}).
  
\bibitem[{\citenamefont{Park et~al.}(2008)\citenamefont{Park, Son, Yang, Cohen,
  and Louie}}]{park:nal08}
\bibinfo{author}{\bibfnamefont{C.-H.}~\bibnamefont{Park}},
  \bibinfo{author}{\bibfnamefont{Y.-W.} \bibnamefont{Son}},
  \bibinfo{author}{\bibfnamefont{L.}~\bibnamefont{Yang}},
  \bibinfo{author}{\bibfnamefont{M.~L.} \bibnamefont{Cohen}}, \bibnamefont{and}
  \bibinfo{author}{\bibfnamefont{S.~G.} \bibnamefont{Louie}},
  \bibinfo{journal}{Nano Lett.} \textbf{\bibinfo{volume}{8}},
  \bibinfo{pages}{2920} (\bibinfo{year}{2008}).

\bibitem[{\citenamefont{Brey and Fertig}(2009)}]{brey:prl09}
\bibinfo{author}{\bibfnamefont{L.}~\bibnamefont{Brey}} \bibnamefont{and}
  \bibinfo{author}{\bibfnamefont{H.~A.} \bibnamefont{Fertig}},
  \bibinfo{journal}{Phys. Rev. Lett.} \textbf{\bibinfo{volume}{103}},
  \bibinfo{pages}{046809} (\bibinfo{year}{2009}).

\bibitem[{\citenamefont{Park et~al.}(2009)\citenamefont{Park, Son, Yang, Cohen,
  and Louie}}]{park:prl09}
\bibinfo{author}{\bibfnamefont{C.-H.} \bibnamefont{Park}},
  \bibinfo{author}{\bibfnamefont{Y.-W.} \bibnamefont{Son}},
  \bibinfo{author}{\bibfnamefont{L.}~\bibnamefont{Yang}},
  \bibinfo{author}{\bibfnamefont{M.~L.} \bibnamefont{Cohen}}, \bibnamefont{and}
  \bibinfo{author}{\bibfnamefont{S.~G.} \bibnamefont{Louie}},
  \bibinfo{journal}{Phys. Rev. Lett.} \textbf{\bibinfo{volume}{103}},
  \bibinfo{pages}{046808} (\bibinfo{year}{2009}).

\bibitem[{\citenamefont{J.~M. Pereira~Jr. and
  Vasilopoulos}(2007)}]{pereira:apl07}
\bibinfo{author}{\bibnamefont{J.~M. Pereira~Jr.}},
\bibinfo{author}{\bibfnamefont{P.}~\bibnamefont{Vasilopoulos}}
  \bibnamefont{and}
  \bibinfo{author}{\bibfnamefont{F.~M.}~\bibnamefont{Peeters}},
  \bibinfo{journal}{Appl. Phys. Lett.} \textbf{\bibinfo{volume}{90}},
  \bibinfo{pages}{132122} (\bibinfo{year}{2007}).

\bibitem[{\citenamefont{Barbier et~al.}(2008)\citenamefont{Barbier, Peeters,
  Vasilopoulos, and Pereira}}]{barbier:prb08}
\bibinfo{author}{\bibfnamefont{M.}~\bibnamefont{Barbier}},
  \bibinfo{author}{\bibfnamefont{F.~M.} \bibnamefont{Peeters}},
  \bibinfo{author}{\bibfnamefont{P.}~\bibnamefont{Vasilopoulos}},
  \bibnamefont{and} \bibinfo{author}{\bibnamefont{J.~M. Pereira~Jr.}},
  \bibinfo{journal}{Phys. Rev. B}
  \textbf{\bibinfo{volume}{77}}, \bibinfo{pages}{115446}
  (\bibinfo{year}{2008}).

\bibitem[{\citenamefont{Barbier et~al.}(2009)\citenamefont{Barbier,
  Vasilopoulos, and Peeters}}]{barbier:prb09}
\bibinfo{author}{\bibfnamefont{M.}~\bibnamefont{Barbier}},
  \bibinfo{author}{\bibfnamefont{P.}~\bibnamefont{Vasilopoulos}},
  \bibnamefont{and} \bibinfo{author}{\bibfnamefont{F.~M.}
  \bibnamefont{Peeters}}, \bibinfo{journal}{Phys. Rev. B}
  \textbf{\bibinfo{volume}{80}}, \bibinfo{pages}{205415}
  (\bibinfo{year}{2009}).

\bibitem[{\citenamefont{Bliokh et~al.}(2009)\citenamefont{Bliokh, Freilikher,
  Savel'ev, and Nori}}]{bliokh:prb09}
\bibinfo{author}{\bibfnamefont{Y.~P.} \bibnamefont{Bliokh}},
  \bibinfo{author}{\bibfnamefont{V.}~\bibnamefont{Freilikher}},
  \bibinfo{author}{\bibfnamefont{S.}~\bibnamefont{Savel'ev}}, \bibnamefont{and}
  \bibinfo{author}{\bibfnamefont{F.}~\bibnamefont{Nori}},
  \bibinfo{journal}{Phys. Rev. B} \textbf{\bibinfo{volume}{79}},
  \bibinfo{pages}{075123} (\bibinfo{year}{2009}).

\bibitem[{\citenamefont{Arovas et~al.}()\citenamefont{Arovas, Brey, Fertig,
  Kim, and Ziegler}}]{arovas:10}
\bibinfo{author}{\bibfnamefont{D.~P.} \bibnamefont{Arovas}},
  \bibinfo{author}{\bibfnamefont{L.}~\bibnamefont{Brey}},
  \bibinfo{author}{\bibfnamefont{H.~A.} \bibnamefont{Fertig}},
  \bibinfo{author}{\bibfnamefont{E.-A.} \bibnamefont{Kim}}, \bibnamefont{and}
  \bibinfo{author}{\bibfnamefont{K.}~\bibnamefont{Ziegler}},
  \bibinfo{journal}{New J. Phys.} \textbf{\bibinfo{volume}{12}},
  \bibinfo{pages}{123020} (\bibinfo{year}{2010}).

\bibitem[{\citenamefont{Barraza-Lopez et~al.}(2010)\citenamefont{Barraza-Lopez, Vanevic,
  Kindermann, and Chou}}]{barraza:prl10}
\bibinfo{author}{\bibfnamefont{S.}~\bibnamefont{Barraza-Lopez}},
  \bibinfo{author}{\bibfnamefont{M.}~\bibnamefont{Vanevi\'c}},
  \bibinfo{author}{\bibfnamefont{M.}~\bibnamefont{Kindermann}},
  \bibnamefont{and} \bibinfo{author}{\bibfnamefont{M.~Y.}~\bibnamefont{Chou}},
  \bibinfo{journal}{Phys. Rev. Lett.} \textbf{\bibinfo{volume}{104}},
  \bibinfo{eid}{076807} (\bibinfo{year}{2010}).

\bibitem[{\citenamefont{B\"uttiker}(1990)}]{buettiker:prl90}
\bibinfo{author}{\bibfnamefont{M.}~\bibnamefont{B\"uttiker}},
  \bibinfo{journal}{Phys. Rev. Lett.} \textbf{\bibinfo{volume}{65}},
  \bibinfo{pages}{2901} (\bibinfo{year}{1990}).

 \bibitem[{\citenamefont{Jalabert}(1992)}]{jalabert94}
\bibinfo{author}{\bibfnamefont{ R.A. } \bibnamefont{  Jalabert}}
  \bibnamefont{and}
  \bibinfo{author}{\bibfnamefont{ J.-L.}~\bibnamefont{Pichard}}  \bibnamefont{and}\bibinfo{author}{\bibfnamefont{C.~W.~J.} \bibnamefont{Beenakker}},
  \bibinfo{journal}{Europhys. Lett.} \textbf{\bibinfo{volume}{27}},
  \bibinfo{pages}{255} (\bibinfo{year}{1994}).
  
\bibitem[{\citenamefont{Cheianov et~al.}(2009)\citenamefont{Cheianov, and Fal'ko}}]{falko:prb06}
\bibinfo{author}{\bibfnamefont{V.~V.} \bibnamefont{Cheianov}},
  \bibinfo{author}{\bibfnamefont{V.~I.}~\bibnamefont{Fal'ko}},
  \bibinfo{journal}{Phys. Rev. B} \textbf{\bibinfo{volume}{74}},
  \bibinfo{pages}{041403(R)} (\bibinfo{year}{2006}).



\end{thebibliography}
\end{document}